\documentclass[11pt]{article}
\usepackage{amsmath}
\usepackage{graphicx}
\usepackage{amssymb}
\usepackage{amsfonts}
\usepackage{latexsym}
\usepackage{psfig}

\def\beq{\begin{eqnarray}}
\def\eeq{\end{eqnarray}}

\def\al{\alpha}
\def\be{\beta}

\def\ga{\gamma}\def\de{\delta}
\def\vp{\varepsilon}

\def\ka{\kappa}
\def\la{\lambda}
\def\na{\nabla}

\def\rh{\rho}
\def\si{\sigma}

\def\th{\theta}

\def\La{\Lambda}

\begin{document}

\title{On the torsion effects of a relativistic spin fluid in early cosmology}
\author{G. de Berredo-Peixoto $\mbox{}^{1}$ \footnote{Email: 
gbpeixoto@hotmail.com} \quad and E.A. de Freitas $\mbox{}^{1\,,2}$ \footnote{Email: 
emanuel@fisica.ufjf.br} \\ \\
1. Departamento de F\'\i sica, ICE, Universidade Federal de Juiz de Fora \\
Campus Universit\' ario - Juiz de Fora, MG Brazil  36036-330 \\ \\
2. Col\' egio T\' ecnico Universit\' ario, Universidade Federal 
de Juiz de Fora \\
Av Bernardo Mascarenhas, 1283 - Juiz de Fora, MG Brazil 36080-001}
\date{}
\maketitle

\begin{quotation}
\noindent
{\large{\bf Abstract.}}
In this work we investigate the effects of torsion in 
the framework of Einstein-Cartan theory in early
cosmology. We study solutions for a homogeneous 
and isotropic relativistic 
Weyssenhoff spin fluid 
with dynamical timelike axial current, 
also homogeneous and isotropic.
The general solutions can mostly be 
described by means of three particular
solutions. The properties of these solutions
(such as singularity avoidance and primordial 
or late accelerated expansion)
are analysed and depend on the relations
between the source parameters.

\vskip 1mm


\end{quotation}
\newpage

\section{Introduction}

The effects of non-Riemannian structures in the framework of Cosmology
have been studied long ago. The most simple generalization of General
Relativity (preserving metricity) is achieved by the introduction of 
an asymetric connection, with torsion as its antisymmetric part. The Riemannian
case is obtained by imposing zero torsion. It is possible to consider
cosmological models in the framework of more general non-Riemannian
structures, such as non-metricity (see for example the review by Puetzfeld 
\cite{Puetzfeld}), but these cases will not be treated here. There are a large number 
of papers dealing with torsion in several approaches, with wide applications. 
For the introduction of the foundations of the theory of gravity with torsion, 
see Ref. \cite{revhehl}, and for a more recent review, including the quantum 
aspects of torsion, see Ref. \cite{shapiro}. 

In the seventies, it was discovered that
the singularity avoidance and inflation can be induced by torsion 
in the Einstein-Cartan theory, by Kopczynski \cite{kopcz},
Trautman \cite{trautman} and Hehl {\it et al} \cite{hehl}. 
In the Einstein-Cartan theory, torsion is not dynamical and
is completely expressed in terms of the spin sources \cite{revhehl}. 
Thus, in order to study the effects of torsion in the Einstein-Cartan 
theory, one has to introduce matter with spin. This can be done in 
several ways. One of the most natural ways is to consider, besides the
Einstein-Cartan action, the Lagrangian describing spin-$1/2$ Dirac 
fields, minimally interacting with torsion. It is possible to describe
this theory as a modified General Relativity with a spin-spin 
contact interaction. According to Ref. \cite{kerlick}, torsion does not 
prevent the initial singularity, but rather enhances it. In addition, 
torsion can provide accelerated expansion phase in a metric-torsion
theory with matter described by Dirac and Rarita-Schwinger 
spinors \cite{watanabe}. In this scenario, the authors of Ref. 
\cite{kremer1} described the transition between early accelerated
expansion and decelerated one in terms of massive Dirac fields (see also, for example, Ref. \cite{kremer2}). It is possible of course
to consider another Lagrangian for gravity and torsion, in
a more general context of Poincar\'e Gauge Theory of Gravity, 
with quadratic terms in curvature strenghs. Within this possibility,
one can mention for example the works in Refs. \cite{minke1,minke2}
(see also references therein), which take into account a spinless
matter.
 
Another way to introduce torsion is to consider a fluid with intrinsic 
spin density, which in principle does not admit a Lagrangian description based
on spinor fields. One
has to postulate a spin correction to the energy-momentum tensor. For
exemple, Szydlowski and Krawiec \cite{krawiec} have studied the cosmological
effects of an exotic perfect fluid\footnote{This fluid presents an intrinsic spin density (See also Ref. \cite{1983} where a quantum treatment was applied
for this fluid in early cosmology).} known as the Weyssenhoff
fluid \cite{weyssenhoff}, as well
as the constraints from supernovae Ia type observations, concluding that 
the dust Weyssenhoff fluid provides accelerated expansion 
but it can not serve as an
alternative to Dark Energy. Also, in Ref. \cite{Puetzfeld2}, Puetzfeld and Chen
derived some constraints from supernovae Ia data in a different scenario of
non-Riemannian geometry. Gasperini \cite{gasperini} considered the
Weyssenhoff fluid with its energy momentum tensor (derivable from a 
Lagrangian) previoulsy improved by Ray and Smalley \cite{ray}, with spin as a thermodynamical variable. 
In Ref. \cite{gasperini}, torsion provides 
singularity avoidance and early accelerated expansion, 
but the expansion factor of the 
cosmological scale, $a(t)$, is too small, unless the state equation parameter $w$ ($p = w \rho$) of the spin fluid 
is fine tunned in a very special way. See also Ref. \cite{lasenby} where similar 
analysis was performed in a more general context without assuming a particular
metric. Obukhov and Korotky \cite{obukhov} formulated 
a more general variational theory describing the Weyssenhoff fluid and
also applied to cosmological models with rotation, shear and 
expansion. Recently, B\"ohmer and Burnett \cite{bohmer1} introduced
a special spinorial matter satisfying the Cosmological Principle
\cite{tsamparlis} (see also \cite{bohmer2}), which, together with
interaction terms and without cosmological constant, 
mimics vacuum energy responsible for inflation.

It is worth mentioning that many works without torsion but with
different matter sources (besides the scalar field) 
inducing inflation can be 
found in literature. For example, Golounev, Mukhanov and Vanchurin
\cite{mukhanov} proposed a scenario with massive non-minimally 
coupled vector field, which induces inflation. In homogeneous and 
isotropic universe, these vector fields behave as a 
minimally coupled massive field, and they can be introduced as
an orthogonal triplet or in a large number randomly oriented in
order to provide isotropic inflation. There are no clear understanding
about the physical motivations for these vector fields. The authors 
of Ref. \cite{mota1} consider also time-like non-minimally coupled vector fields (see also \cite{mota2} where cosmological perturbations were studied). Among theories without torsion, one can mention also
Ref. \cite{novello}, where the authors show that the dark energy 
can be described by means of the usual electrodynamics with a 
non-linear adding term. Also, 
one can consider a timelike vector field, responsible for 
violation of Lorentz symmetry \cite{carroll}. In the present 
work, the timelike axial current not only violates Lorentz symmetry,
but it is related to torsion in the context of the Einstein-Cartan
theory, and it is originated by the Weyssenhoff fluid. 
On the other side, we
can cite Ref. \cite{mota3}, where the possibility of inflation 
induced by non-standard spinors was investigated, as well as 
their imprints on CMB anisotropies.

We consider the Einstein-Cartan theory with both the spin-spin contact interaction and the Weyssenhoff fluid. In practice, one has an axial 
current and a spin density as additional sources besides the usual 
perfect fluid energy density and pressure. In early universe, any matter
content has very high temperature, so we let $w$ fixed corresponding to
a radiation fluid ($w = 1/3$), which mimics the ultra-relativistic regime. 
If the model is supposed to describe early inflation, then the spin fluid
is an exotic form of matter which can play the role of vacuum energy. 
The matter content of the model is given by this exotic fluid (with $w=1/3$)
plus a timelike axial current, homogeneous and isotropic.
The anisotropic case is very interesting (providing description of 
primordial anisotropy), although it is beyond the framework of the 
present work.

This paper is organized as follows. In section 2, 
we shall present a brief
introduction to Einstein-Cartan theory, including 
the dynamical equations of the model. In section 3, 
we investigate the solutions of the model for different cases.
In section 4, we draw our conclusions and final remarks.

\section{Einstein-Cartan Theory and dynamical equations}

The action in the Einstein-Cartan framework is given by
\beq
S = \int \sqrt{-g} d^4 x\left\{ -\frac{1}{\ka^2} (\tilde{R} - 2\Lambda )+ 
{\cal L}_M\right\}\, , \label{general action}
\eeq
where metric has signature ($+ - - -$), $\ka^2 = 16 \pi G$ (we use units
such that $\hbar = c = 1$), $\Lambda$ is the cosmological constant 
and $\tilde{R}$ is the Ricci scalar
constructed with the asymmetric connection\footnote{All quantities with 
an upper tilde are constructed with the asymmetric connection,
and the corresponding quantities without tilde are constructed with
the Riemannian (symmetric) conection.}, 
$\tilde{\Gamma}^\mu\mbox{}_{\al\be}$, which, by using the metricity
condition ($\tilde{\na}_\al g_{\mu\nu} = 0$) and the following definition of torsion, 
$$
T^\mu\mbox{}_{\al\be} := \tilde{\Gamma}^\mu\mbox{}_{\al\be} - 
\tilde{\Gamma}^\mu\mbox{}_{\be\al}\, ,
$$
can be expressed as 
\beq
\tilde{\Gamma}^\mu\mbox{}_{\al\be} = 
\Gamma^\mu\mbox{}_{\al\be} + K^\mu\mbox{}_{\al\be}\, ,
\eeq
where $\Gamma^\mu\mbox{}_{\al\be}$ is the Riemannian connection
(Levi-Civita connection) and
the quantity $K^\mu\mbox{}_{\al\be}$ is the contortion tensor,
given by
$$
K^\mu\mbox{}_{\al\be} = \frac{1}{2}\left( T^\mu\mbox{}_{\al\be}
-T_\al\mbox{}^\mu\mbox{}_\be - T_\be\mbox{}^\mu\mbox{}_\al 
\right)\, .
$$
The term ${\cal L}_M$ is the Lagrangian describing matter distribution. 
We consider here the following matter Lagrangian:
\beq
{\cal L}_M = {\cal L}_{AC} + {\cal L}_{SF}\, ,
\eeq
where ${\cal L}_{SF}$ is the Lagrangian of the spin fluid \cite{ray}
and ${\cal L}_{AC}$ is the external source, present in the minimally 
coupling Dirac sector (see, e.g., \cite{shapiro}):
\beq
{\cal L}_{AC} = J^\mu S_\mu\, .
\eeq 
Here, $S_\mu$
is the axial part of torsion, defined by $S_\mu = \varepsilon_{\la\rho\si\mu}
T^{\la\rho\si}$ ($\varepsilon_{\la\rho\si\mu}$ is 
the Levi-Civita tensor,
with $\varepsilon_{0123} = \sqrt{-g}$), and
$J^\mu$ is the external axial current\footnote{The matrix 
$\ga^5$ is the chiral Dirac matrix $\ga^5 = (i/4!)
 \varepsilon^{\al\be\mu\nu}\ga_\al\ga_\be\ga_\mu\ga_\nu = 
i\ga_0\ga_1\ga_2\ga_3$.}, 
$J^\mu =\, < \bar{\psi}\ga^5\ga^\mu\psi\,>$, where this average
is due to quantum effects (see \cite{shapiro}), such that $J^\mu$ 
is a vacuum property, responsible for Lorentz violation (see also
\cite{kostelecky}).

In order to vary the action and get the dynamical equations, one has
to define what are the independent variables. We choose $g^{\mu\nu}$
and $T^\al\mbox{}_{\be\ga}$ as independent dynamical variables, and $J^\mu$ 
as a quantity defined by the symmetry violation of the vacuum, 
which is, as we shall see, dependent 
from dynamical variable $g^{\mu\nu}$. The spacetime metric 
is the spatially flat homogeneous and isotropic metric such that
\beq
ds^2 = dt^2 - a(t)^2 (dx^2+dy^2+dz^2)\,.
\label{metric}
\eeq
It is natural to assume that, in comoving frame, $J^\mu$ is a 
homogeneous and isotropic vector, otherwise it would break
the isotropy of the universe. Thus, $J^\mu$ is a timelike vector
so that we let $J^\mu J_\mu = J^2(t)$. 

\subsection{Variational principle and dynamical equations}

The dynamical equations for metric fields and torsion, in terms of the
sources, can be obtained respectively by the usual 
procedure\footnote{We use similar notations from literature, 
e.g., Ref. \cite{gasperini}.}
\beq
\frac{\de S}{\de g^{\mu\nu}} = 0 & \Longrightarrow &
\frac{1}{\sqrt{-g}}\frac{\de (\sqrt{-g}\tilde{R})}{\de g^{\mu\nu}} = 
\frac{\ka^2}{2} T_{\mu\nu} \\
\frac{\de S}{\de T^\mu\mbox{}_{\nu\al}} = 0 & \Longrightarrow &
\frac{1}{\sqrt{-g}}\frac{\de (\sqrt{-g}\tilde{R})}
{\de T^\mu\mbox{}_{\nu\al}} =
\ka^2 \theta_\mu\mbox{}^{\nu\al}\, , \label{theta}
\eeq
where
$$
T_{\mu\nu} := \frac{2}{\sqrt{-g}}\frac{\de (\sqrt{-g}
({\cal L}_{AC} + {\cal L}_{SF}))}{\de g^{\mu\nu}} :=
T^{AC}_{\mu\nu} + T^{SF}_{\mu\nu}
$$
and
$$
\theta_\mu\mbox{}^{\nu\al} := \frac{1}{\sqrt{-g}}\frac{\de (\sqrt{-g}
({\cal L}_{AC} + {\cal L}_{SF}))}{\de T^\mu\mbox{}_{\nu\al}} := (\theta_{AC})_\mu\mbox{}^{\nu\al} + 
(\theta_{SF})_\mu\mbox{}^{\nu\al}\, .
$$
It should be noticed that instead of procedure 
(\ref{theta}), one can verify that 
\beq
\frac{\de S}{\de K_{\mu\nu\al}} = 0 \;\;\; {\rm gives}\;\;\;
\frac{1}{\sqrt{-g}}\frac{\de (\sqrt{-g}\tilde{R})}{\de K_{\mu\nu\al}} =
\ka^2 \tau^{\mu\nu\al}\,, \label{K}
\eeq
where
$$
\tau^{\mu\nu\al} := \frac{1}{\sqrt{-g}}\frac{\de (\sqrt{-g}
{\cal L}_M)}{\de K_{\mu\nu\al}} := \tau^{\mu\nu\al}_{AC} + \tau^{\mu\nu\al}_{SF}\,.
$$
Equation ({\ref{K}) is totally equivalent to equation (\ref{theta}),
thus we shall use it for the convenient correspondence with
notations in literature.

By the convention for the curvature tensor in the form\footnote{
We use the symbols $[\,]$ and $(\,)$ to denote antisymmetrization
and symmetrization, according to
$$
A_{[\mu\nu ]} = \frac{1}{2}(A_{\mu\nu} - A_{\nu\mu}) \;\;\; 
{\rm and }\;\;\;
A_{(\mu\nu )} = \frac{1}{2}(A_{\mu\nu} + A_{\nu\mu})\, .
$$}
$\tilde{R}^\mu\mbox{}_{\la\al\be} = 
2\tilde{\Gamma}^\mu\mbox{}_{\la [\be ,\al ]}  + 
\tilde{\Gamma}^\rho\mbox{}_{\la\be} \tilde{\Gamma}^\mu\mbox{}_{\rho\al}
- \tilde{\Gamma}^\rho\mbox{}_{\la\al} \tilde{\Gamma}^\mu\mbox{}_{\rho\be}$,
one can achieve, disregarding total derivatives, the relation
$$
\int d^4x \sqrt{-g}\tilde{R} = \int d^4x \sqrt{-g}\left( 
R + K^\al\mbox{}_{\rho\al} K^{\rho\la}\mbox{}_\la - 
K^\al\mbox{}_{\rho\la} K^{\rho\la}\mbox{}_\al\right)
$$
Variation with respect to $T^\mu\mbox{}_{\al\be}$ gives
\beq
T^{\mu\al\be} + 2g^{\mu [\al}T^{\be]} = \ka^2 \tau^{\be\al\mu}\, ,
\label{eqK}
\eeq
where $T^\be = T^{\rho\be}\mbox{}_\rho$.
In order to express $\tau^{\be\al\mu}$, let us mention that the
contribution from spin fluid is given by  
$\tau^{\be\al\mu}_{SF} = -\frac{1}{2} S^{\be\al} u^\mu $ \cite{ray},
where $S^{\be\al}$ is the spin tensor (antissymmetric), and
$u^\mu$ is the fluid four-velocity\footnote{In previous papers, 
like, e.g., Ref. \cite{gasperini}, the expression
$\tau^{\be\al\mu}_{SF}$ has the opposite sign. The reason is
that we adopt $L_{SF}$ with different sign in order to
reproduce the same fluid dynamical equations usual in literature.}.
By straightforward algebra,
we can obtain the expression for $\tau^{\be\al\mu} = 
\tau_{AC}^{\be\al\mu} + \tau_{SF}^{\be\al\mu}$ and 
consequently,
\beq
T^{\mu\al\be} + 2g^{\mu [\al}T^{\be]} = 
\ka^2\left\{
-\frac{1}{2} S^{\be\al} u^\mu + 
2\varepsilon^{\be\al\mu\rho} J_\rho \right\}\,.
\eeq
Using the Weyssenhoff condition\footnote{Also known as the
Frenkel condition. It is included by hand, but emerges 
automatically in the formalism proposed by Obukhov and Korotky
\cite{obukhov}.}, $S^{\be\al} u_\al = 0$, 
from the above equation we can derive
\beq
T^{\mu\al\be} = -\ka^2\left\{
2\varepsilon^{\mu\al\be\rho} J_\rho -
\frac12 S^{\al\be} u^{\mu} \right\}\, 
\label{source1}
\eeq
and
\beq
S_\si = \varepsilon_{\mu\al\be\si} (2K^{\mu\al\be}) = 
12\ka^2 J_\si + 
\frac{1}{2}\ka^2\varepsilon_{\mu\al\be\si} S^{\mu\al} u^\be\, .
\label{source2}
\eeq

Variation with respect to $g^{\mu\nu}$ can be done in 
a straightforward way, giving
\beq
G_{\mu\nu} - g_{\mu\nu}\Lambda & - & 
K^\al\mbox{}_{\mu\al}K^\la\mbox{}_{\nu\la}-
\frac{1}{2}T^\al\mbox{}_{\rho\mu}T^\rho\mbox{}_{\nu\al}
- \frac{1}{2}T^\al\mbox{}_{\mu\la}T_\al\mbox{}^\la\mbox{}_\nu
-\frac{1}{4}T_{\mu\rho\la}T_\nu\mbox{}^{\rho\la} \nonumber \\
& + & 
\frac{1}{8}g_{\mu\nu}\left(
4K^\al\mbox{}_{\rho\al}K^{\la\rho}\mbox{}_\la +
2T^{\al\rho\la}T_{\rho\la\al} - T^{\al\rho\la}T_{\al\rho\la}\right)
= \frac{\ka^2}{2}T_{\mu\nu}\, .
\label{eqg}
\eeq
To write the dynamical equations, one can substitute (\ref{source1})
and (\ref{source2}) into the appropriate quantities in (\ref{eqg}),
including $T_{\mu\nu}$. In this way we can rewrite equation 
(\ref{eqg}) in the form
\beq
G_{\mu\nu} - g_{\mu\nu}\Lambda & - & 
\ka^4\left\{
g_{\mu\nu} J^2 + 2 J_\mu J_\nu -
\frac14 g_{\mu\nu} J_\si \varepsilon^{\al\rho\la\si} S_{\rho\la} u_\al -
\frac12 J_\si \varepsilon_{(\mu}\mbox{}^{\rho\la\si} u_{\nu )} S_{\rho\la}
\right. \nonumber \\
& + & \left.
\frac{1}{32} g_{\mu\nu} S_{\rho\la}S^{\rho\la} - \frac18 S_{\mu\la}S_\nu\mbox{}^\la
+\frac{1}{16} u_\mu u_\nu S_{\rho\la}S^{\rho\la}\right\} 
= \frac{\ka^2}{2}(T^{AC}_{\mu\nu} + T^{SF}_{\mu\nu})\, ,
\label{eqg2}
\eeq
where $J^2 = J_\si J^\si$.
Notice that from (\ref{source1}) one has $T^\al\mbox{}_{\mu\al} = 0$, such that 
several terms in (\ref{eqg}) vanish. Thus, with axial current and the spin fluid
satisfying the Frenkel condition, $S^{\be\al} u_\al = 0$, there are only traceless
degrees of freedom of torsion.

The next step is averaging the above equation. 
For this purpose, a natural and 
simple assumption is $<S_{\al\be}> = 0$, which means that
although the spin tensor might have a particular 
direction in the microscopic scale, 
its mean value vanishes at macroscopic domain (i.e., 
the particles have a randomic spin distribution). 
Let us define (see Ref. \cite{gasperini})
\beq
< S_{\al\be} S^{\al\be} > = 2 \si^2 \, ,
\eeq 
such that 
$$
< S_\mu\mbox{}^\la S_{\nu\la} > = 
\frac{2}{3}(g_{\mu\nu} - u_\mu u_\nu ) \si^2\, .
$$

Now we have to express $T^{AC}_{\mu\nu}$ and $T^{SF}_{\mu\nu}$
in terms of the sources. For $T^{SF}_{\mu\nu}$, one can obtain
the formula (see, e.g., Ref. \cite{gasperini}):
\beq
T^{SF}_{\al\be} & = & u_{(\al} S_{\be )}\mbox{}^\mu u^\nu 
K^\rho\mbox{}_{\mu\nu} u_\rho +
u^\rho K^\mu\mbox{}_{\si\rho} u^\si u_{(\al} S_{\be )\mu}
-\frac{1}{2} u_{(\al} T_{\be )\mu\nu} S^{\mu\nu} 
\nonumber \\ & + & 
\frac{1}{2} T_{\nu\mu (\al} S^\mu\mbox{}_{\be )} u^\nu
+ 2\ka^2\left\{ (\rho + p)u_\al u_\be - 
p g_{\al\be}\right\}\,. \label{gasper}
\eeq
Substituting (\ref{source1}) into (\ref{gasper}), one
obtains
\beq
T^{SF}_{\al\be} & = &
\ka^2 J^\rho u_{(\al} \vp_{\be )\mu\nu\rho} S^{\mu\nu} -
\ka^2 J^\rho S^\mu\mbox{}_{(\be}\vp_{\al )\mu\nu\rho} u^\nu -
\frac{\ka^2}{4}u_\al u_\be S_{\mu\nu} S^{\mu\nu} +
\frac{\ka^2}{4} S_{\mu\al} S^\mu\mbox{}_\be \nonumber \\
& + &
2\ka^2 \left\{(\rho + p)u_\al u_\be - pg_{\al\be}\right\}\,.
\eeq
We achieve, by averaging,
\beq
< T^{SF}_{\al\be} > & = &
\ka^2 < J^\rho u_{(\al} \vp_{\be )\mu\nu\rho} S^{\mu\nu} > -\,
\ka^2 < J^\rho S^\mu\mbox{}_{(\be}\vp_{\al )\mu\nu\rho} u^\nu >
\nonumber \\
& - &\frac{2\ka^2}{3} u_\al u_\be \si^2 + 
\frac{\ka^2}{6} g_{\al\be}\si^2 +
2\ka^2 \left\{(\rho + p)u_\al u_\be - pg_{\al\be}\right\}\,.
\eeq
It should be mentioned that, although $< S_{\mu\nu} >$
vanishes, it is possible to take $< S_{\mu\nu} J^\al > \neq 0$. 
Nevertheless, we argue that the assumption $< S_{\mu\nu} J^\al > = 0$ 
is correct because  $S_{\mu\nu}$ is randomic only in the 3-space,
and $J^\mu$ has not any spatial component.

Let us finally express $T^{AC}_{\mu\nu}$ in terms of the sources.
The variation of $\sqrt{-g}{\cal L}_{AC}$ with respect to 
$g^{\mu\nu}$ should be done with special care, since $J^\mu$ (and 
also $J_\mu$) depends on $g^{\mu\nu}$. This computation is done
in the Appendix. The result can be written as
\beq
T^{AC}_{\mu\nu}  = 
-2\vp_{(\mu}\mbox{}^{\al\be\rho} T_{\nu )\al\be} J_\rho - 
S_{(\mu} J_{\nu )}\,.
\label{tAC}
\eeq
Substituting the sources (\ref{source1}) and (\ref{source2}),
we have
\beq
T^{AC}_{\mu\nu}  = 
-8\ka^2 g_{\mu\nu} J^2 - 4\ka^2 J_\mu J_\nu -
\ka^2 \vp_{(\mu}\mbox{}^{\be\rho\la} u_{\nu )} S_{\be\rho} J_\la
+\frac{\ka^2}{2}\vp_{(\mu}\mbox{}^{\be\rho\la} 
J_{\nu )} S_{\rho\la} u_\be\,.
\eeq  

Now, one can obtain the following dynamical equation after 
averaging (using $< S_{\mu\nu} J^\al > = 0$):
\beq
G_{\mu\nu} = \ka^4 \left\{
- 3g_{\mu\nu} J^2 + \frac{1}{16}g_{\mu\nu} \si^2 - 
\frac{1}{8} u_\mu u_\nu \si^2 \right\}
 +  \frac{\ka^2}{2} \left\{(\rho + p) u_\mu u_\nu - 
p g_{\mu\nu}\right\} + \Lambda g_{\mu\nu}\, ,
\label{eqdyn}
\eeq

\subsubsection*{Dynamical equations}

Let us consider the fluid in a relativistic regime, 
such that $p = \rho/3$.
For the metric (\ref{metric}), the relevant components of equation 
(\ref{eqdyn}) can be written as
\beq
\frac{3\dot{a}^2}{a^2} & = & \ka^4\left\{
-3J^2 - \frac{\si^2}{16}\right\} + \frac{\ka^2}{2} \,\rho
+ \La \label{eqdyn1} \\
-\frac{\dot{a}^2}{a^2} - \frac{2\ddot{a}}{a} & = & 
\ka^4\left\{ 3J^2 - \frac{\si^2}{16}\right\} +
\frac{\ka^2}{6} \,\rho - \La\, ,
\label{eqdyn2}
\eeq
where the dot above variables means time derivative. From 
these equations, one can get
\beq
\frac{\ddot{a}}{a} = \ka^4\left\{
-J^2 + \frac{\si^2}{24}\right\} - 
\frac{\ka^2}{6}\rho + \Lambda/3\, .
\label{eqdynddot}
\eeq
It is remarkable that the axial current enters in the above
equation with different sign from the spin contribution. In 
this equation, the axial current acts as a kind of density 
energy of some ordinary matter. We see that only the 
cosmological constant and the spin contribute to
the accelerated expansion. 

The energy conservation law
can be obtained by comparing equation (\ref{eqdyn2}) with the 
time derivative of equation (\ref{eqdyn1}):
\beq
\dot{\rho} a + 4\dot{a}\rho  = \ka^2\left\{
\frac{1}{8a^5}\frac{d}{dt}\left(a^6 \si^2 \right) + 
6 a \frac{d}{dt}(J^2)
\right\}\, .
\label{energyconserv}
\eeq

Let us remark that the above equation describes energy conservation
in the general case when all fields interact with each other. Of course,
the spin contribution ($\si^2$) always interacts with the fluid
itself, because spin is a fluid property. On the other hand, we have
freedom to choose an interacting axial current and a non-interacting
one. These possibilities define two classes of solutions. 
The non-interacting axial current case was studied previoulsy in 
Ref. \cite{ourpaper}. The case without spinning fluid can be found 
in Ref. \cite{shapiro}, where a different variational procedure
was realized.

Equations (\ref{eqdyn1}) and (\ref{energyconserv}), along with 
the initial conditions $\rho(t_0) = \rho_0$ and $a(t_0) = a_0 = 1$
($t_0$ is the present time), determine the dynamical solution of
the model. For the CMB, we know that 
$\rho_0\sim\rho_c \Omega_{{\rm rad}} $. As
$$
\Omega_{{\rm rad}}\sim 10^{-5} \;\;\; {\rm and} \;\;\;
\rho_c = \frac{3H^2}{8\pi G}\sim 4 \times 10^{-47} 
(GeV)^4\;\; ({\rm with}\;\; H = \frac{\dot{a}}{a})\, ,
$$
one can estimate $\rho_0\sim 10^{-52} (GeV)^4$ 
for pure radiation. Of course,
this value does not necessarily correspond to the exotic spin fluid, 
but can serve as a reference (or upper bound). In what follows, we 
shall consider in most cases
a much lower density for the spinning fluid, say, 
$\rho_0\sim 10^{-54} (GeV)^4$.
  
\section{Solutions with interacting axial current}

In searching for solutions, we should specify the dependence
of $J^2$ and $\si^2$ on the density, $\rho$. One can adopt 
$\si^2 = \ga \rho^{3/2}$ ($\ga =$ positive constant) \cite{ponomariev}.
We identify the axial current as coming from the spinning fluid.
In doing so, the natural choice for $J^2$ is 
$J^2 = \be \rho^{3/2}$ ($\be =$ positive constant). With these 
assumptions, the dynamical equations for the model are
\beq
\dot{\rho} = \frac{12\ka^2\dot{a}\ga\rho^{3/2} - 64\dot{a}\rho
}{16a - 3\ka^2 a\ga\rho^{1/2} - 144\ka^2a\be\rho^{1/2}}\,  
\label{rho}
\eeq
and
\beq
\dot{a} = a \sqrt{\frac{\ka^2}{6}\rho + \frac{\La}{3} - 
\ka^4\rho^{3/2}\left(\be + \frac{1}{48}\ga\right)}\,.
\label{a}
\eeq

The above system is very complicated and can not be solved 
analytically. However, we can extract relevant information 
just from (\ref{rho}). Let us rewrite it:
\beq
\frac{d\rho}{da} = \frac{12\ka^2\ga\rho^{3/2} - 64\rho
}{16a - 3\ka^2 a\ga\rho^{1/2} - 144\ka^2a\be\rho^{1/2}}\,.
\label{rho2}
\eeq
For a particular $\rho = \rho_{f}$, there is a fixed point, 
i.e., $d\rho /da = 0$. When $\rho(a)$ reaches $\rho_f$, it
ceases to vary. One can express $\rho_f$ as
$$
\rho_f = \frac{256}{9\ka^4\ga^2}\,.
$$
Also, there is an apparent singularity when $\rho = \rho_{c}$
such that
$d\rho /da \to \infty$. The value for $\rho_{c}$ is obtained 
by straighforward algebra:
$$
\rho_{c} = \frac{256}{(3\ka^2\ga + 144\ka^2\be)^2}\,.
$$
Actually, the system is not only apparently singular at $\rho_c$,
but it is really singular, because
$\dot{a}$ (from (\ref{a})) can not vanish for $\rho_c$.
Notice that always $\rho_{c} < \rho_{f}$. Let us investigate 
the three unique possibilities: 
(i) $\rho_0 < \rho_{c} < \rho_{f}$; 
(ii) $\rho_{c} < \rho_0 \leq \rho_{f}$ and
(iii) $\rho_{c} < \rho_{f} \leq \rho_0$.
For the first possibility, we have $\rho^{\prime} < 0$ in the
whole interval $\rho < \rho_{c}$. 
Thus, as time goes backward,
density is increased until it reaches $\rho_c$. It means
that in some finite $a > 0$ in the past, $\rho^{\prime}$ is infinite.
As it has not clear meaning, we must reject the possibility (i).

For the second possibility, we have $\rho^{\prime} > 0$ in the
whole interval $\rho_{c} < \rho < \rho_{f}$. Using similar reasoning
as before, $\rho^{\prime}$ becomes again infinite for some finite
$a$ in the past. In the last possibility, $\rho^{\prime} < 0$
in the interval $\rho_{f} \leq \rho$, such that $\rho$ never
reaches $\rho_c$. Thus, the range given under (iii) appears to 
be the only viable physical choice.

By the above considerations, the following inequality must
be satisfied:
$$
\frac{256}{9\ka^4\ga^2} \leq \rho_0\,.
$$
In numbers, it means that the dimensionless parameter $\ga$ has a
lower bound determined by $\rho_0$. For $\rho_0 = 10^{-54}$ 
GeV$^4$, we achieve a very large lower bound\footnote{Notice that
$\La = 5\times 10^{-84}$ GeV$^2$ and 
$\ka^2 = 3.38\times 10^{-37}$ GeV$^{-2}$.}: 
$\ga \geq 1.58\times 10^{64}$! This lower bound imposes the lower 
bound $\si^2(t_0) = \ga \rho_0^{3/2} \geq 1.6\times 10^{-17}$
GeV$^6$.

However, a carefull analysis shows that the assumption
$\si^2\,,J^2 \propto \rho^{3/2}$ is not rigorous. Indeed,
one should start from a more simple and fundamental assumption,
by considering the dependence of $\si^2$ and $J^2$
(or $(\bar{\psi}\psi)^2$) on the scale factor, as
\beq
\si^2\,,J^2 \propto a^{-6}\, .
\label{ansatz}
\eeq
Notice that in the previous paper \cite{gasperini}
the fluid density satisfies $\rho \propto a^{-4}$,
such that $\si^2 \propto \rho^{3/2}$. Nevertheless, 
$\rho \propto a^{-4}$ is clearly not an exact solution
of the equation (\ref{energyconserv}). Thus, 
the ansatz (\ref{ansatz}) will be adopted from
now on. Let then
\beq
J^2 = \frac{J^2_0}{a^6}\, 
\;\;\; {\rm and} \;\;\;
\si^2 = \frac{\si^2_0}{a^6}\, .
\eeq
The equations of motion (\ref{eqdyn1}) and 
(\ref{energyconserv}) can be written in the form
\beq
\frac{\dot{a}^2}{a^2} = 
-\kappa^4\left( \frac{J^2_0}{a^6} + 
\frac{\si^2_0}{48 a^6}\right)
+\frac{\kappa^2}{6}\rho + \frac{\Lambda}{3}\,,
\label{eq1}
\eeq
\beq
\frac{d\rho}{da} = -\frac{4\rho}{a} - 
\frac{36\kappa^2 J^2_0}{a^7}\, .
\label{eq2}
\eeq
It is quite difficult to solve analytically the above system, so
it is convenient to treat it numerically. The first observation 
is that the quantity $d\rho/da$ in equation (\ref{eq2})
is negative and lower than $-4\rho /a$, 
thus the effect of
the axial torsion is that, as far as $\rho_0$ ($=\rho (t_0)$)  
is taken to be the same quantity for both cases 
(with and without axial current), the values for $\rho$ 
are higher for $t < t_0$.

Now, let us consider equation (\ref{eq1}). Its right hand 
side must be positive for all values of $a(t)$. Choosing 
the special case $a = 1$, we arrive at the following condition
(for $\rh_0 = 10^{-54}$ GeV$^4$):
\beq
J^2_0 + \frac{\si^2_0}{48} < \frac{\rho_0}{6\kappa^2}
+ \frac{\Lambda}{3\kappa^4} \approx 
4.38\times 10^{-11}{\rm GeV}^6\, ,
\label{ubound}
\eeq
which defines an upper bound constraint for the source 
parameters $J^2_0$ and $\si^2_0$. In equation (\ref{eq1}),
the term proportional to $\ka^4$ has the same dependence on
the scale factor than the non-Riemannian parameters
in Ref. \cite{Puetzfeld2}. Although a different model is
studied in this work, this coincidence opens the possibility
for estimating some constraints from supernovae Ia data
applied to the present model\footnote{We thank the anonymous
referee for this indication.}. As a result, we get 
$J^2_0 + \frac{\si^2_0}{48} 
\leq 5.91\times 10^{-12} {\rm GeV}^6$, which is remarkably
similar with (\ref{ubound}).

Before investigating general solutions for (\ref{eq1}) and
(\ref{eq2}), let us consider the obvious particular
solution of (\ref{eq2}):
\beq
\rho (a) = \frac{\rho_0}{a^6}\,.
\eeq 
Substituting this solution into (\ref{eq2}), one achieves 
the constraint
\beq
\rho_0 = 18\kappa^2 J^2_0\, .
\label{constr}
\eeq
Notice that in this case the equation (\ref{ubound})
establishes effectively an upper bound for $\si^2_0$, 
since $J^2_0$ is attached to $\rho_0$.
The equation (\ref{eq1}) reads
\beq
\frac{\dot{a}^2}{a^2} = \frac{\th}{a^6}
+ \frac{\Lambda}{3}\, \;\;\; {\rm where} \;\;\;
\th := \kappa^4
\left(2J^2_0 - \frac{\si^2_0}{48}\right)\,.
\eeq
There are three solutions depending on the
sign of $\th$: 
\beq
a(t) = \left\{
\sqrt{\frac{3\th}{\Lambda}} 
\sinh \,(\sqrt{3\Lambda}\, t)\right\}^{1/3}
\;\;\; {\rm for} \;\;\; \th > 0\, ,
\label{sol1}
\eeq
where the integration constant was already fixed by $a(0) = 0$,
and the second solution is given in the implicit form:
\beq
a(t)^3 + \sqrt{\frac{3\th}{\Lambda} + a(t)^6} = 
\sqrt{-\frac{3\th}{\Lambda}} \exp(\sqrt{3\Lambda}\,t)
\;\;\; {\rm for} \;\;\; \th < 0\, .
\label{sol2}
\eeq
In the above solution, the integration constant was choosen
such that $a(0) = a_{{\rm min}} = (-3\th / \Lambda )^{1/6}$, 
with $a_{{\rm min}}$ being the minimum value of $a$ which 
can be found by the condition $\th + \Lambda a^6 \geq 0$. 
Thus, for $\th < 0$, there is a singularity avoidance, 
and the universe undergoes an accelerated expansion
all the time. In contrast, solution (\ref{sol1}) does not
prevent the model from having initial singularity.

The third solution comes from the case $\th = 0$. It
is given by $a(t)\propto e^{\sqrt{\Lambda /3}\,t}$ (De Sitter). 
It is remarkable that in this case the Universe expands
as it was empty with only the cosmological constant, but
there is a fluid with $\rho\propto a^{-6}$.

It is interesting to extract information about the source
parameters, $J^2_0$ and $\si^2_0$, just from the experimental
constraints such as the known age of the universe\footnote{Notice
that this proceedure is not rigorous if we remember that in
late times, the role of torsion will be supressed by the
conventional matter content, such as perfect fluid in dust form.}, 
$t_0 = 13.7$ billion years $= 6.56\times 10^{41}$ GeV$^{-1}$.
Let us consider, for example, the solution (\ref{sol1}).
By using the expression for $a(t)$ in (\ref{sol1}) at
the equality $a(t_0) = 1$, one can get $\th = 
4.18\times 10^{-86}$ GeV$^{2}$ (similar quantity can be found
for $|\th|$ in the case (\ref{sol2})). Now, using equation
(\ref{eqdynddot}) in the form
$$
\frac{\ddot{a}}{a} = -\frac{2\th}{a^6} + \frac{\Lambda}{3}\, ,
$$
we achieve, by direct substitution,
$$
\frac{\ddot{a}}{a} = \frac{\Lambda}{3}\left\{
1 - {\rm csch}^2\,(\sqrt{3\Lambda} t)\right\}\,.
$$
It is possible to show that this quantity is positive
if $t > 2.96\times 10^{41}$ GeV$^{-1}$, what means that,
taking $\th = 4.18\times 10^{-86}$ GeV$^{2}$, the
expansion of the universe is accelerated for
$a > 0.61$. 

Notice that in the absence of the cosmological constant, the 
quantity $\th$ must be positive, admitting then the solution 
(\ref{sol1}). Only a non-null cosmological constant can provide singularity avoidance (for $J^2_0\neq 0$). It is
remarkable that the theory with axial current and spin fluid,
satisfying (\ref{constr}), requires the introduction of $\Lambda$,
otherwise the model would have no solutions for all possible
values of $\si^2_0$. For $J^2_0 = 0$, the model reduces to the
one studied by Gasperini \cite{gasperini}, with early
and late accelerated expansion, and also singularity avoidance.
In this case, the early accelerated expansion takes place in
a very short period of time, and it should be mentioned that 
$J^2_0\neq 0$ can be taken as an important generalization,
in a Lorentz violating theories or as a vacuum quantum effect,
for example.

In fact, condition (\ref{constr}) seems to be quite particular.
However, as we shall see below, the general 
solutions for $\rho (t_0) \neq 18\kappa^2 J^2_0$ can be described
by the particular solutions dictated by $\rho \propto a^{-6}$.

\subsection*{General Solutions}

According to previous considerations, we know that 
$\rho \propto a^{-6}$ is a particular solution, which
demands (\ref{constr}). Indeed, one can perform numerical
integration of equation (\ref{eq2}), 
using the {\it Mathematica} software,
starting from the point $\rho(a_0 = 1) = \rho_0$ satisfying
(\ref{constr}). The integrated curve for $\rho(a)$ will
be exactly the curve for $\rho \propto a^{-6}$, which
can be drawn as a straigh line with negative slope ($-6$)
in the logarithm scaling. 

\begin{figure}[h]
$$
\begin{array}{l}
\psfig{file=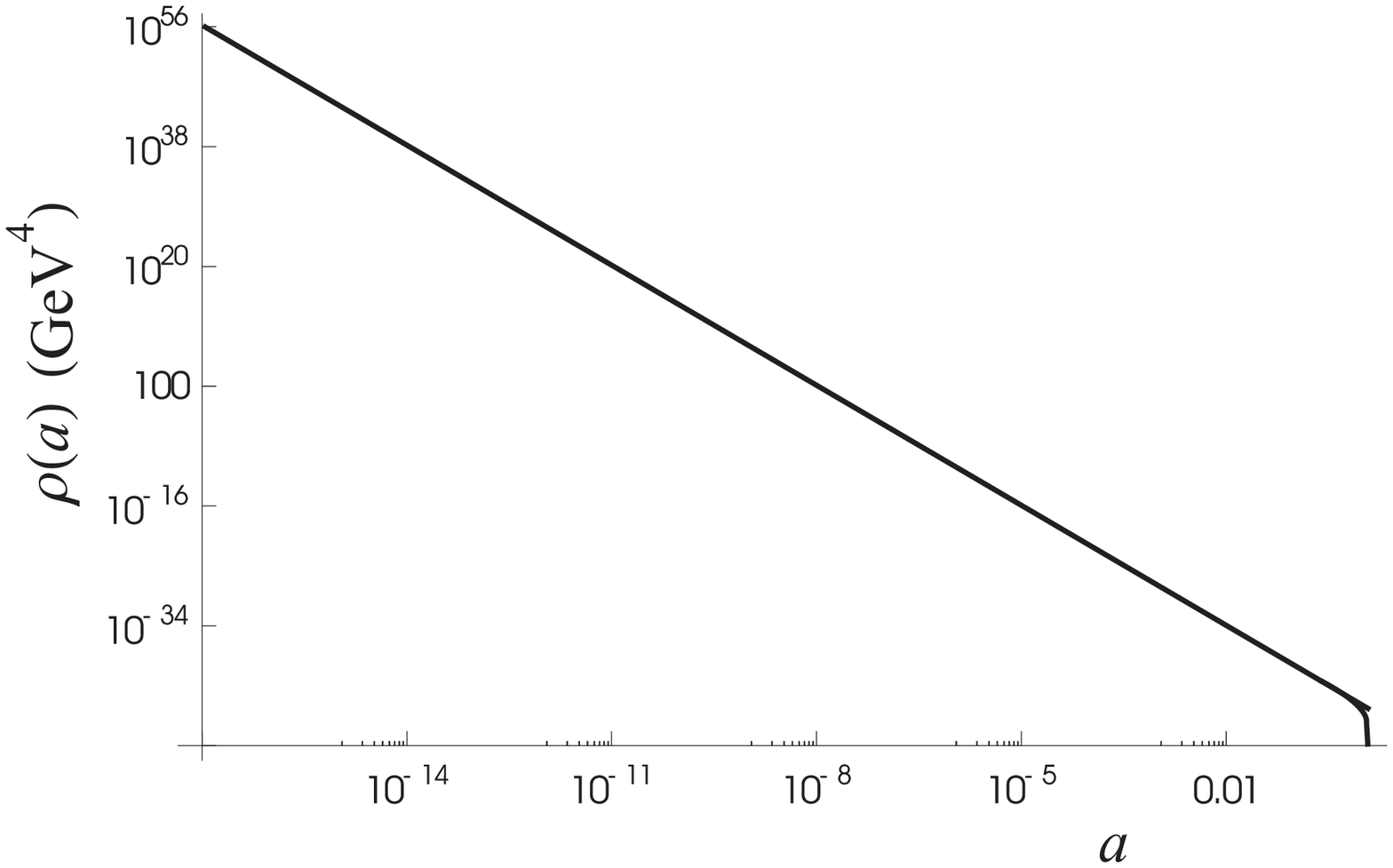,width=7cm,height=4.5cm}\;
\psfig{file=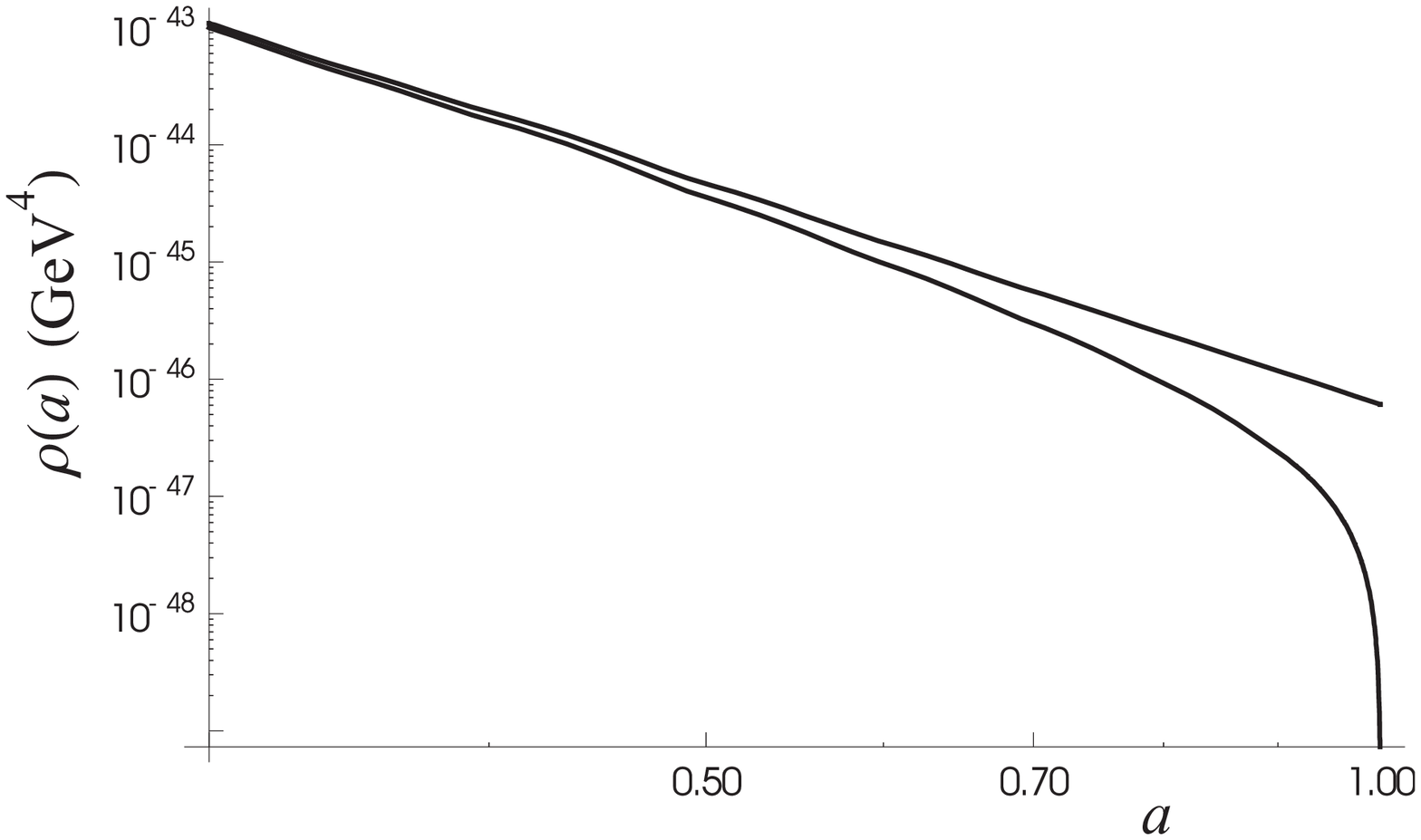,width=7cm,height=4.5cm}
\end{array}
$$
\caption{The integral curves, in logarithm scale, 
for $\rho(a)$, from initial conditions
$\rho (1) = 10^{-54}$ GeV$^4$ and $\rho (1) = 
18 \kappa^2 J^2_0 = 6.08\times 10^{-47}$ GeV$^4$ coincide,
except in the small region shown in the right plot.} 
\end{figure}

In order to numerically integrate, we choose 
some $J^2_0$ compatible with condition (\ref{ubound}),
say, $J^2_0 = 10^{-11}$ GeV$^6$, and $\rh_0$ to be
different from (\ref{constr}). Let us choose a lower
value: $\rh_0 = 10^{-54}$ GeV$^4$. The integration
is shown in the left side of Figure 1, 
together with the integration
obeying (\ref{constr}), corresponding to 
$\rho \propto a^{-6}$. It is remarkable that both
curves coincide, but it is essential to stress that
they coincide in a certain (and wide) interval.
The same plotting in the right side of Figure 1 
is drawn in an interval much more
closer to $a=1$, where the difference between the
two solutions becomes clear.

It seems that this behavior is universal: the 
curve $\rh(a)$, for any $\rho_0$, differs from 
the solution $\rho \propto a^{-6}$ in some region
close to $a = 1$, but in the remaining region, say, 
$0 < a < 0.1$, the solution is very close to 
$\rh(a) = 18 \kappa^2 J^2_0/a^6$. Now, remind that
there is no much room for big values of $\rh_0$, and
we are mostly considering early cosmology, thus the
particular solutions discussed previously are
quite general and instructive. 

We should mention that the case $J^2_0 = 0$ 
\cite{gasperini} is 
completely different, because $\rho(a)$ will
be substantially affected, such that $\rh\propto a^{-4}$.

\section{Conclusions and Discussions}

We have investigated the cosmological effects of the axial 
current together with the relativistic spin fluid ($ p = \rho/3$) 
in Einstein-Cartan theory. As already known 
from literature \cite{kopcz,trautman,hehl}, torsion provides singularity 
avoidance and accelerated expansion. The contribution from 
axial current, however, favours a decelerated 
expansion, in contrast to the spin fluid.

There are two classes of solutions: 
one with an external non-interacting 
axial current, and other with interacting
axial current (i.e., time-dependent axial current).
The first one was considered in previous works 
(see \cite{shapiro} for the case without spin fluid 
and with a conformal global axial vector, and 
\cite{ourpaper} with a global constant vacuum axial vector 
and a spin fluid).

In the present work, the axial current is assumed as a
composite field, $J^\mu = < \bar{\psi}\ga^5\ga^\mu\psi >$,
where the Dirac fields (presumably) describe the fluid itself.
The axial current interacts with the spinning fluid, as 
realized by the energy-momentum conservation. It is essential
that this feature determines the dependence of energy density
on the scale factor, $a$, which is substantially different
from the case with non-interacting axial current.    
It is natural to assume that both spin and axial current decay 
in similar way as the universe expands.

We conclude that, based on the ansatz $J^2\propto a^{-6}$ and
$\si^2\propto a^{-6}$, the general solutions behave
(in a relevant domain) as the particular solutions coming 
from $\rho\propto a^{-6}$, such that their properties 
are the same. These properties were analised in details
for the particular solutions: If the source parameter
$\th$ is positive, the universe has an initial singularity 
and its expansion is decelerated ($\ddot{a} < 0$) until
some epoch, when late accelerated expansion begins.
This epoch depends basically on the parameter $\th$, which
can be determined from the known age of universe. By
doing this, late accelerated expansion starts at $a = 0.61$.
In both particular solutions, $J^2_0$ can be fixed by the
present value of density, $\rho_0$. Thus, $\si^2_0$ 
determines the sign of $\th$.

For the case $\th < 0$, universe has an accelerated 
expansion all the time, and the solution features 
singularity avoidance. In this case, equation
(\ref{ubound}) represents an upper bound for 
the quantities $J^2_0$ and $\si^2_0$. 

As discussed above, the general solutions have a 
remarkable behavior (as the shift shown in the right side of 
Figure 1) in the region close to the present day. 
This is very strange in the physical point of view, 
because it would be an enormous coincidence if
the shift of the actual curve occurs right on the
present time, $t_0$. Thus, it seems that the physically
reasonable solution must be the particular one, 
$\rh\propto a^{-6}$. As a consequence, the parameter
$J^2_0$ should be related to the present density
by equation (\ref{constr}).

\subsubsection*{Acknowledgments}
The work of the authors has been supported by 
research grant from CNPq (G.B.P), from FAPEMIG (G.B.P. and E.A.F.) and 
FAPES (G.B.P.). We would like to express our gratitute
to Prof. Ilya Shapiro for stimulating discussions and for relevant 
suggestions on reading the manuscript.

\appendix

\subsection*{Appendix: variational proceedure}
 
Here we shall calculate the functional derivative of $J^\mu$. To
do so, one must take into account that 
$J^\mu = \bar{\psi}\ga^5\ga^\mu\psi$, with 
$\ga^5 = (i/4!)\vp^{\al\be\mu\nu}\ga_\al\ga_\be\ga_\mu\ga_\nu$.
As $\ga_\mu\ga^\mu = 4$, we can write 
$\de (\ga_\al\ga^\al)/\de g^{\mu\nu} = 0$. With this equation,
one can express $\de \ga_\al /\de g^{\mu\nu}$ in terms of
$\de \ga^\be /\de g^{\mu\nu}$ and vice-versa. Using 
$\ga^\rho = g^{\rho\la}\ga_\la$, we achieve
\beq
\frac{\de \ga^\rho}{\de g^{\mu\nu}} = 
\frac{1}{2}\de^\rho_{(\mu}\ga_{\nu )}
\;\;\;\; {\rm and} \;\;\;\;
\frac{\de \ga_\rho}{\de g^{\mu\nu}} = 
-\frac{1}{2}g_{\rho(\mu}\ga_{\nu )}\,.
\label{app1}
\eeq

Now, we know that $\vp^{\al\be\rho\la} = E^{\al\be\rho\la}/\sqrt{-g}$,
where $E^{\al\be\rho\la}$ is the Levi-Civita symbol, which is clearly
independent on the metric. Thus, we obtain
\beq
\frac{\de \vp^{\al\be\rho\la}}{\de g^{\mu\nu}} = 
\frac{1}{2}g_{\mu\nu}\vp^{\al\be\rho\la}
\;\;\;\; {\rm and} \;\;\;\;
\frac{\de \vp_{\al\be\rho\la}}{\de g^{\mu\nu}} = 
-\frac{1}{2}g_{\mu\nu}\vp_{\al\be\rho\la}\, .
\label{app2}
\eeq
Using (\ref{app1}) and (\ref{app2}), one can get 
$$
\frac{\de \ga^5}{\de g^{\mu\nu}} = 0\, .
$$
With all these results, one obtains by straighforward algebra
the variation of $J^\al$ and $J_\al$:
\beq
\frac{\de J^\rho}{\de g^{\mu\nu}} = 
\frac{1}{2}\de^\rho_{(\mu}J_{\nu )}\,,
\;\;\;\;\;\;
\frac{\de J_\rho}{\de g^{\mu\nu}} = 
-\frac{1}{2}\de^\rho_{(\mu}J_{\nu )}
\;\;\; {\rm and} \;\;\;
\frac{\de J^2}{\de g^{\mu\nu}} = 0\,.
\eeq

Similar computations can be performed for 
$S^\la = \vp^{\al\be\rho\la}T_{\al\be\rho}
= \vp^{\al\be\rho\la} g_{\al\si} T^\si\mbox{}_{\be\rho}$.
After all, the variation of $\sqrt{-g}J^\mu S_\mu$ can
be expressed by means of result (\ref{tAC}).

\end{document}